\documentclass[11pt,aps,pra]{revtex4}
\usepackage{graphicx}
\def\fr{\frac}
\def\a{\alpha}
\def\b{\beta}

\def\g{\gamma}
\def\k{\kappa}
\def\l{\lambda}

\def\r{\rho}

\def\G{\Gamma}

\def\ra{\rangle}
\def\la{\langle}
\def\be{\begin{equation}}
\def\ee{\end{equation}}
\def\bea{\begin{eqnarray}}
\def\eea{\end{eqnarray}}
\begin{document}
\title{Coherent states for exactly solvable potentials}
\author{T. Shreecharan$^{\mathrm{1,2}}$\footnote{e-mail: shreet@prl.ernet.in},
Prasanta K. Panigrahi$^{\mathrm{1,2}}$\footnote{e-mail:
prasanta@prl.ernet.in} and J.
Banerji$^{\mathrm{1}}$\footnote{e-mail: jay@prl.ernet.in}}

\affiliation{$^{\mathrm{1}}$Physical Research
Laboratory, Navrangpura, Ahmedabad-380 009, India\\
$^{\mathrm{2}}$School of Physics, University of Hyderabad,
Hyderabad-500 046, India}
\begin{abstract}
A general algebraic procedure for constructing coherent states of
a wide class of exactly solvable potentials e.g., Morse and
P{\"o}schl-Teller, is given. The method, {\it a priori}, is
potential independent and connects with earlier developed ones,
including the oscillator based approaches for coherent states and
their generalizations. This approach can be straightforwardly
extended to construct more general coherent states for the quantum
mechanical potential problems, like the nonlinear coherent states
for the oscillators. The time evolution properties of some of
these coherent states, show revival and fractional revival, as
manifested in the autocorrelation functions, as well as, in the
quantum carpet structures.
\end{abstract}
\maketitle

\section{Introduction}

Coherent states (CS) and their generalizations, in the context of
harmonic oscillators, are well-studied in the literature
\cite{glauber,klauder2,perelomov,gilmore}. Algebraic approaches
have been particularly useful for providing a unified treatment of
these states and their inter-relationships. For example, in Ref.
\cite{shanta}, not only a general procedure for constructing a
large class of oscillator based CS has been provided, but it has
also been shown that, some of these states are dual to each other
in a well-defined manner. The algebraic approaches
straightforwardly lead to the construction of squeezed and other
states showing interesting non-classical features. These elegant
and powerful algebraic procedures of construction owe their
origin, partly, to the simplicity of the Heisenberg-Weyl algebra;
$[a,a^{\dagger}]=1$, characterizing the harmonic oscillator. Based
on the symmetries and keeping in mind the desired requirements,
various procedures have been developed for constructing CS for
Morse \cite{benedict,roy,fakhri,nietoprd3}, Hydrogen atom
\cite{gerry}, P{\"o}schl-Teller
\cite{nietoprd2,crawford,klauder3,kinani} and other potentials
\cite{mohanty}. The role of Morse potential in molecular physics
is well-known \cite{child}. The study of CS for hydrogenic atoms
have assumed increasing importance in light of their relevance to
Rydberg states \cite{stroud}, which may find potential application
for quantum information processing \cite{rangan}.

It is known that, all the criteria desired of a coherent state and
found in the oscillator based CS, e.g., minimum uncertainty
product, eigenstate of the annihilation operator (AO) and
displacement operator states are not simultaneously achievable in
other potential based CS. Hence, a number of CS, diluting one or
more of the above criteria, have been constructed in the
literature.

In the context of algebraic approaches, supersymmetric (SUSY)
quantum mechanics \cite{khare} based raising and lowering
operators have found significant application. In particular,
eigenstates of the lowering operator for Morse \cite{benedict} and
P{\"o}schl-Teller \cite{kinani} potentials have been found and
their properties studied. Recently Antoine et. al. \cite{klauder3}
have constructed Klauder type CS for the P{\"o}schl-Teller
potential, using a matrix realization of ladder operators, their
motivation being the temporal stability of the CS. It is
well-known that the SUSY ladder operators act on the Hilbert space
of different Hamiltonians except for the case of the harmonic
oscillator. Establishing a precise connection between the complete
set of states, describing the above CS, and the symmetries of
these potential problems has faced difficulties
\cite{fakhri,jellal}. To be specific, in case of the
Barut-Girardello CS for the Morse potential, the ladder operators
are taken to be functions of quantum numbers, which has led to
problems in defining a proper algebraic structure. To resolve the
same, some authors \cite{fakhri} have resorted to the introduction
of additional angular coordinates \cite{asim}.

In this work, we provide an algebraic construction of the CS for a
wide class of potentials, belonging to the confluent
hypergeometric (CHG) and hypergeometric (HG) classes. The
procedure is based on a simple method of solving linear
differential equations (DEs) \cite{guru1}, which enables one to
express the solutions in terms of monomials. In the space of
monomials, it is straightforward to identify various types of
ladder operators, their underlying algebraic structures
\cite{guru2}, and construct lowering operator eigenstate in a
transparent manner. The fact that, the monomials and the quantum
mechanical eigenfunctions are connected through similarity
transformations, enables one to preserve these algebraic
structures at the level of the wavefunctions and simultaneously
obtain the desired CS. Thus the coherent state is initially
potential independent. The information about a specific potential
is then incorporated by fixing the parameters of the series and
also the ground state wavefunction of the potential under study.
The known results for CS are obtained in specific limits. In
addition, our procedure demonstrates the construction of more
general CS, similar to the nonlinear CS \cite{sudarshan} in the
oscillator example. The origin of confusion in identifying the
algebraic structure in SUSY based approaches is then pointed out
and subsequently resolved in a natural manner. It is shown that
the recently found CS for various potentials are related to the
two different realizations of SU(1,1) algebra.

The paper is organized as follows. Keeping in mind the connection
of the wavefunctions of the exactly solvable potentials, to be
considered here, and the solutions of the CHG and HG DEs, we first
briefly outline the relationship between the solutions of the
above equations and the space of monomials. This connection is
then made use of, to identify the ladder operators and the
symmetry algebras in the space of monomials. AOCS are constructed,
through a recently proposed procedure for solving linear DEs. This
method yields the precise connection between the present, and the
earlier oscillator based approaches for constructing CS. In Sec.
III, the potential independent CS, thus constructed, are connected
with Morse and P{\"o}schl-Teller potentials, as examples. The
advantage of constructing the ladder operators in the space of
monomials is then pointed out by resolving the difficulties in
identifying the symmetry generators in the SUSY based approach.
Various earlier known CS are then derived as special cases. Time
evolution properties of some of these CS are then studied.
Autocorrelation functions and the underlying quantum carpet
structures \cite{frank} of these CS clearly reveal the phenomena
of revival and fractional revival \cite{banerji}. We conclude in
Sec. IV after pointing out a number of open interesting problems,
where the present procedure can be profitably employed. Keeping in
mind the rich symmetry structure of the hydrogen atom problem and
the various procedures employed for constructing the corresponding
CS, we desist from the analysis of these CS here; this will be
taken up in a future project.

\section{Algebraic construction of coherent states}

This section is devoted to the construction of CS in a manner
which is potential independent. We make essential use of the fact
that the exactly solvable potentials, to be studied here, belong
to the CHG and the HG classes, whose solutions can be connected to
the space of monomials through similarity transformations. In the
space of monomials, the identification of symmetry algebras and
their ladder operators become easy. The AO eigenstates in the
space of monomials are first found through the above mentioned
procedure of solving DEs and then connected to those at the level
of the polynomials, through similarity transformation. Connection
with earlier oscillator based approaches is also exhibited.

A single variable linear DE, of arbitrary order, can be cast in
the form $[F(D)+P(x,d/dx)]y(x)=0$; where $F(D)$ is a function of
the Euler operator $D=x d/dx$ and $P(x,d/dx)$ contains rest of the
operators. A formal series solution of the above DE is given by
\cite{guru1,guru2}
 \be \label{an}
 y(x)= C_\a \sum_{n=0}^{\infty}(-1)^n
 \left[\fr{1}{F(D)}P(x,d/dx)\right]^n x^\a \, ,
 \ee
provided the condition $F(D)x^\a=0$, is satisfied. This condition
fixes the value of $\a$. Here $C_\a$ is the normalization
constant. The proof is straightforward and follows by direct
substitution. In the CHG and the HG cases, $\a=n$ (integer)
corresponds to polynomial solutions and for $\a \neq n$ series
solutions are obtained.

Following an earlier work \cite{guru2}, the simplest set of
operators, at the level of monomials, which give rise to a SU(1,1)
algebra can be written as \cite{brif}
 \be \label{one}
 K_+ = x,K_- = \left[x\fr{d^2}{dx^2}+ b\fr{d}{dx}\right],\,{\rm and}\,K_3 =
 \left[x\frac{d}{dx}+\frac{b}{2}\right].
 \ee
Here $b$ is deliberately taken same as the CHGDE parameter, so
that the algebra is well defined at the level of the
wavefunctions. It should be noted that, these symmetry algebras
are not unique, and the interesting consequences arising from this
fact will be pointed out later. The AOCS corresponding to $K_-$
satisfies,
 \be \label{def1}
K_- \varphi(x,\b) = -\b \varphi(x,\b) \,.
 \ee
Left multiplying by $x$:
 \be
\left(x^2\frac{d^2}{dx^2} + b x\frac{d}{dx} + \b x \right)
\varphi(x,\b) = 0 \,
 \ee
one can identify $F(D)=(D+b-1)D$ and $P(x,d/dx)=\b x$. The
condition $F(D)x^\a=0$ gives $\a=0,1-b$. For $\a=0$ one obtains,
 \begin{widetext}
 \bea \nonumber
 \varphi(x,\b)&=& N(\b)^{-1}
\sum_{n=0}^{\infty}
 (-1)^n\left[\fr{\b}{(D+b-1)D}x\right]^n x^0\,,\\
 &=& N(\b)^{-1} \sum_{n=0}^{\infty}\fr{(-1)^n}{n!}\fr{(\b
 x)^n}{(b)_n}\, = \exp\left[-\frac{\b}{(D+b-1)}x\right]\,.
 \eea
 \end{widetext}
$N(\b)^{-1}$ is the normalization constant. The commutation
relation, $[D,x^d]=d x^d$, and the definition of the Pochhammer's
symbol, $(b+n-1) \cdots b=\G(b+n)/\G(b)=(b)_n$, have been used in
the above derivation. We can make immediate contact with the
oscillator based approach of Ref. \cite{shanta}. Denoting
 \be
\tilde{K}_+ \equiv \frac{1}{(D+\b-1)}x\,,
 \ee
it is found that, they obey the Heisengerg-Weyl algebra,
$[K_-,\tilde{K}_+]=1$. Such conjugate operators were explicitly
constructed in \cite{shanta} and used to obtain a large class of
multi-photon coherent states.

Knowing that the solution of the CHGDE can be written in the form
\cite{guru1}
 \be \label{chsol}
 \Phi(-n;b;x) = (-1)^{n}
 \frac{\G(b)}{\G(b+n)} e^{-K_-} x^{n}\,,
 \ee
one can obtain the coherent state at the level of the polynomial
 and later on at the level of the wavefunctions through appropriate similarity
 transformations, as illustrated below.
 Denoting $S \equiv \exp(-K_-)$, one obtains $S K_- S^{-1} S \varphi(x,\b) = \b
S \varphi(x,\b)=\b \tilde{\varphi}(x,\b)$, where
$\tilde{\varphi}(x,\b)$ is the coherent state at the level of the
polynomial:
 \be \label{chgcs}
 \tilde{\varphi}(x,\b) =
N(\b)^{-1}\sum_{n=0}^{\infty} \fr{\b^n}{n!} \Phi(-n;b;x)\, .
 \ee
It will be seen explicitly in the subsequent section that
introducing appropriate measures through similarity
transformations one would obtain CS for various quantum mechanical
systems. It is worth emphasizing that the algebraic structures are
transparently preserved in this approach.

Analogously for HG case, taking the lowering operator to be
\cite{guru2}
 \be
 \hat{K}_- = \fr{1}{(D+b)}\left(x\fr{d^2}{dx^2}+c
\fr{d}{dx}\right) \, ,
 \ee
the coherent state at the level of polynomials is given by
 \be \label{hgcs}
 \tilde{\chi}(x,\g) =
N(\g)^{-1}\sum_{n=0}^{\infty} \fr{\g^n}{n!}{_2}F_1(-n,b;c;x) \,,
 \ee
since
 \be \label{hg}
  _2F_1(-n,b;c;x) = (-1)^{n}
\fr{\G(b+n)\G(c)}{\G(c+n)\G(b)}e^{-\hat{K}_-} x^{n} \, .
 \ee
 We can connect to the oscillator based approach in this case
 also; calling
 \be
 \bar{K}_+ \equiv
  \frac{(D+b-1)}{(D+c-1)}x \,,
  \ee
it satisfies the oscillator algebra: $[{\hat K}_-,\bar{K}_+]=1$.
It should be noted that the states obtained using the lowering
operator ${\hat K}_-$ are nonlinear CS \cite{sudarshan}. They are
defined to be eigenstates of the AO of the type $f({\mathcal
N})T_-$, $f({\mathcal N})$ being a function of the number operator
in the oscillator case and the Euler operator in the present one;
$T_-$ is an arbitrary lowering operator. It was recently shown
\cite{wang} that, the AOCS and the Perelomov CS are unified in the
framework of nonlinear CS.

Equations (\ref{chgcs}) and (\ref{hgcs}) reveal that these CS are
quite general; since the eigenfunctions of the exactly solvable
potentials to be considered here, arise as special cases of CHG
and HG series, their corresponding CS will also follow from the
above two general expressions. In fact, starting from the lowering
operator
 \be
 {\tilde T}_- =
 \frac{(D+b_1)\cdots(D+b_p)}{(D+a_1)\cdots(D+a_p)}\frac{d}{dx}\,;
 \ee
more general nonlinear CS can be found:
 \be
 \xi(x,\a)=\sum_{n=0}^{\infty} \frac{\a^n}{n!}
 {_{p+1}}F_p(a_1,\cdots,a_p,-n;b_1,\cdots,b_p;x).
 \ee
Here ${_{p+1}}F_p$ is the generalized HG series. For the sake of
completeness it should be pointed out, the above summation yields
\cite{bateman}
 \be
\xi(x,\a)={_{p}}F_p(a_1,\cdots,a_p;b_1,\cdots,b_p;-x\a) e^\a.
 \ee
It is worth noting that the weight factors associated with the
above CS, play an important role in the study of photon number
statistics in quantum optics, where similar type of states also
appear \cite{appl}.

\section{Connection with potentials}

We now proceed to specific potentials for the purpose of
illustration and establishing connections with the various CS
obtained so far. We use the Morse potential
\cite{benedict,roy,fakhri,nietoprd3} as an example, for the CHG
class of potentials, and the P{\"o}schl-Teller (PT) potential
\cite{nietoprd2,crawford,klauder3,kinani} for the HG class.

The one dimensional Morse potential is given by
 \be
V_{\rm M}= d[1- \exp(- a y)]^2,
 \ee
$a$ and $d$ (depth of the potential) being constant parameters.
Introducing dimensionless parameter $\mu = \sqrt{2md}/a\hbar$,
dimensionless coordinate $q=ay$ and using the transformation rule:
$x=2\mu \exp(-q)$; the Schr{\"o}dinger equation yields the
eigenfunction
 \be
\psi^{\rm M}_n(x) \propto e^{-x/2} x^{\l/2}L_n^{\l}(x) \,,
 \ee
where we have set $\l=2\mu-2n-1$ for the sake of convenience.
Multiplying Eq. (\ref{chgcs}) from the left, by the ground state
of the Morse eigenstate: $\psi^{\rm M}_0(x)\equiv
\exp(-x/2)x^{\l/2}$, and noting that, for $b=\l+1$ the CHG series
can be expressed in terms of the Laguerre polynomials
\cite{stegun}
 \be
 \Phi(-n;\l+1;x)=\fr{n!}{(\l+1)_n}L_n^{\l}(x) \, ;
 \ee
the coherent state for the Morse potential can be written from the
results of the previous section:
 \be \label{laggf1}
 {\tilde \varphi}_{\rm M}(x,\b) = N(\b)^{-1}\G(\l+1)
 \psi^{\rm M}_0\sum_{n=0}^{\infty}\fr{\b^n
 L_n^{\l}(x)}{\G(\l+n+1)}.
 \ee
 After normalization one finds,
 \bea \nonumber
 {\tilde \varphi}_{\rm M}(x,\b) &=&
 \frac{|\b|^{\l/2}}{\sqrt
 {I_{\l}(2|x|)}}\sum_{n=0}^{\infty}\fr{\b^n}{\G(\l+n+1)}\psi^{\rm
 M}_n(x)\\
 &=& \frac{|\b|^{\l/2}}{\sqrt
 {I_{\l}(2|x|)}}(\b)^{-\l/2}e^{\b}e^{-x/2}J_{\l}(2
 \sqrt{x\b})\,.
 \eea
Here, $I_{\l}$ is the modified Bessel function of the first kind
and $J_{\l}$ is the Bessel function of the first kind. The same
expression was obtained in \cite{jellal} using the SUSY based
ladder operators.

In order to see the difficulties associated with properly defining
an algebraic structure using SUSY based ladder operators
\cite{jellal}
 \be
A^{\pm}\psi^{\rm
 M}_{n}(x)=\left[x\frac{d}{dx}\pm\frac{1}{2}(2n+\l+1-x)\right],
 \ee
as alluded to earlier, we explicitly write down the action of the
same on the wavefunctions: $A^{+}\psi^{\rm M}_{n}(x)
=\sqrt{(n+1)(n+\l+1)}\psi^{\rm M}_{n+1}(x)$ and $A^{-}\psi^{\rm
M}_{n}(x)=- \sqrt{n(n+\l)}\psi^{\rm M}_{n}(x)$. In earlier works
without explicitly giving the diagonal operator, its action was
inferred from $[A^{+},A^{-}]\psi^{\rm
 M}_{n}(x)$. However, as was noticed in \cite{fakhri} this approach faces
 problem in constructing the Barut-Girardello CS. To better appreciate the
 difficulties and its resolution, we first identify the corresponding
 ladder operators at the level of monomials through similarity
 transformations, which keeps the algebraic structure intact.
 These operators act as
 \be \label{newlow}
 (K_+ +D-n)x^n = x^{n+1} \quad (K_- +D-n)x^n = n(n+\a)x^{n-1}.
 \ee
It can be noticed that as compared to our ladder operators, at the
level of monomials, the above ones contain an additional operator
$D-n$, which yields zero when acting on monomial $x^n$. It can be
straightforwardly seen that the SUSY based $n$ dependent operators
do not lead to a proper algebra, a difficulty noticed in
\cite{fakhri}. However, the $n$ independent operators lead to the
diagonal operator $K_3 = x d/dx + b/2$; together these form a
closed SU(1,1) algebra. Hence, for the construction of AOCS with a
well defined algebraic structure, it is imperative to use $n$
independent operators. It can be seen that for the ground state,
from which the AOCS are constructed, the above $n$ dependent
operator is absent. Hence the expression for the CS derived
earlier \cite{jellal} and the one obtained here, based on the
SU(1,1) algebra, are identical.

As mentioned earlier the SU(1,1) generators written above are not
unique e.g., the following three generators also form a SU(1,1)
algebra \cite{perelomov}:
\begin{widetext}
 \be
 L_+ = x^2\frac{d}{dx}+2(\l+1)x, \quad L_- = \frac{d}{dx},\quad
 {\rm and}\quad L_3 = x\frac{d}{dx}+\frac{(\l+1)}{2}.
 \ee
 The eigenstate of $L_-$ is given by $\exp(\b x)$. The
 corresponding AOCS, at the level of the wavefunction for the Morse
 potential, can be obtained by making use of the following
 identities,
 \be
 \exp\left[-x\frac{d^2}{dx^2}-(\l+2)\frac{d}{dx}\right]. \frac{d}{dx} =
 \frac{d}{dx}.\exp\left[-x \frac{d^2}{dx^2}-
 (\l+1)\frac{d}{dx}\right]
 \ee
and
 \be
\exp\left[-x \frac{d^2}{dx^2}-
 (\l+1)\frac{d}{dx}\right]\exp(\b x) =
\sum_{n=0}^\infty \b^n L_n^\l(x). \ee
 The resulting coherent state
turns out to be the Perelomov coherent state for this potential
\cite{benedict}:
 \be
{\tilde\varphi}^{\rm PER}_{\rm M}(x,\b) =
\frac{(1-|\b|^2)^{(\l+1)/2}}{\sqrt{\G(\l+1)}}x^{\l/2} \exp\left(-
\frac{x}{2}\frac{1+\b}{1-\b} \right)\,.
 \ee
 \end{widetext}
 Hence, the above SU(1,1) algebra gives the operator, whose
 eigenstate is the Perelomov coherent state; recourse has been taken earlier to
 more complicated nonlinear algebras for this purpose \cite{wang}.

We now derive the CS for the PT class of potentials and
concentrate primarily on the symmetric PT (SPT) and PT potentials.
Plots of the weight factors associated with the CS of the above
mentioned potentials will be given along with the quantum carpet
structure and the auto-correlation figures. The quantum carpet and
the autocorrelation plots, transparently bring out the phenomenon
of revival and fractional revival.

The SPT potential is
 \be
 V_{\rm SPT}(y)= \frac{\hbar^2 \a^2}{2m}\frac{\r(\r-1)}{\cos^2\a y}
 \ee
 and the corresponding eigenvalues and eigenfunctions, in the variable
 $x=\sin\a y$, are
 \begin{widetext}
 \bea \nonumber
 E^{\rm SPT}_n &=& \frac{\hbar^2 \a^2}{2m}(n+\r)^2 , \quad n=0,1,2, \cdots
 \\ \label{mptwave}
 \psi_{n}^{\rm SPT}(x) &=& \left[\frac{\a(n!)(n+\r)\G(\r)\G(2\r)}{\sqrt{\pi}
\G(\r+1/2)\G(n+2\r)}\right]^{1/2}(1-x^2)^{\r/2}C_n^{\r}(x)\,.
 \eea
The Gegenbauer polynomials are related to the HG series via the
relation \cite{stegun}
 \be \label{hggeg}
 {_2}F_1(-n,n+2\r;\r+1/2;z) = \fr{n!}{(2\r)_n}C_n^{\r}(1-2z)\, .
 \ee
Multiplying $(1-x^2)^{\r/2}$ from the left in Eq. (\ref{hgcs}) and
using Eqs. (\ref{mptwave}) and (\ref{hggeg}), the coherent state
for the SPT potential is found to be
 \be \label{mptcs}
 \tilde{\chi}_{\rm SPT}(x,\g) = N(\g)^{-1}\sum_{n=0}^{\infty}
\left[\frac{\G(2\r)\G(\r+1/2)\sqrt{\pi}}
{\a(n!)(n+\r)\G(\r)\G(2\r+n)}\right]^{1/2}\g^n \psi_{n}^{\rm
SPT}(x)\,.
 \ee
The normalization constant, $N(\g)^{-1}$, is
 \bea \nonumber
N(\g)^{-1}=\left[\frac{\a
\G(\r)}{\G(\r+1/2)\G(2\r)\sqrt{\pi}S(|\g|)}\right]^{1/2}
 \eea
where,
 \be
 S(|\g|)= \sum_{n=0}^{\infty}
\frac{|\g|^{2n}}{n!(n+\r)\G(2\r+n)}=\frac{1}{|\g|^{2\r}}
\int^{2|\g|}_0 dx I_{2\r-1}(x) \,.
 \ee
It is worth pointing out that the CS for the SPT potential can be
expressed in terms of the Bessel functions by using a generating
function of the Gegenbauer polynomials \cite{stegun}:
 \be
\tilde{\chi}_{\rm MPT}(x,\g) = N(\g)^{-1} \G(\r+1/2) e^{\g x}
\left(\frac{\g}{2}\right)^{1/2-\r} J_{\r-1/2}(\g\sqrt{1-x^2}). \ee

Similarly we can construct the CS for the PT potential. The PT
potential is
 \be
 V_{\rm PT}(x) = \frac{\hbar^2 \a^2}{2m}\left[\frac{\k(\k-1)}{\sin^2\a x} +
\frac{\r(\r-1)}{\cos^2\a x} \right], \quad \k,\r>1 \,,
 \ee
whose energy eigenvalues and the eigenfunctions are
 \bea \nonumber
E_{n}^{\rm PT} &=& \frac{\hbar^2 \a^2}{2m}(\k+\r+2n)^2, \quad n=0,1,2, \cdots
\\ \label{wave}
\psi_{n}^{\rm PT}(x)&=& C_n (\cos\a x)^\l (\sin\a x)^\k
P_n^{(\k-1/2,\r-1/2)}(1-2\sin^2\a x)
 \eea
Here $C_n$ is the normalization constant and given by
\cite{normalization}
 \be \label{norm}
C_n=\left[\frac{2\a(\k+\r+2n)\G(n+1)\G(\k+\r+n)}
{\G(\k+n+1/2)\G(\r+n+1/2)} \right]^{1/2} \,.
 \ee
 Using Eq. (\ref{wave}) and Eq. (\ref{norm}) in Eq. (\ref{hgcs}) the
AOCS for the P{\"o}schl-Teller potential can be written in the
form
 \be \label{fptcs}
 {\tilde\chi}_{\rm PT}(x,\g) =
N(\g)^{-1}\sum_{n=0}^{\infty}\g^n
\left[\frac{\G(\k+1/2)(\l+1/2)_n}{2\a(\k+1/2)_n(\k+\r+2n)\G(n+1)\G(\k+\r+n)}\right]^{1/2}
\psi_{n}^{\rm PT}(x) \, .
 \ee
In the above equation relation between Jacobi polynomials and HG
series has been used \cite{stegun}. As has been done for the Morse
potential one can also construct Perelomov type coherent state
here.
\end{widetext}

The quadratic nature of the eigenspectra of PT and SPT lead to the
possibility of revival and fractional revival in this quantum
system. Keeping this in mind, we now proceed to study the time
evolution property of the above CS. As expected, these states show
a very rich structure involving revival and fractional revivals.
We give in Fig (1) quantum carpet representing the time evolution
of the above state. The plots for the auto correlation are
provided Figs (3) and (4), which clearly bring out the above
features. Interestingly, there have been some recent proposals to
use the fractional revival for the purpose of factorization of
numbers \cite{mack}. In the above quantum carpet the ridges and
the valleys follow a curved path, unlike the square-well case
where these are straight lines \cite{frank}. We also notice richer
structure arising due to interference. The origin of these
structures in the square-well case have been understood, the
present scenario needs a thorough understanding.


\begin{figure}
\centering
\includegraphics[width=4in]{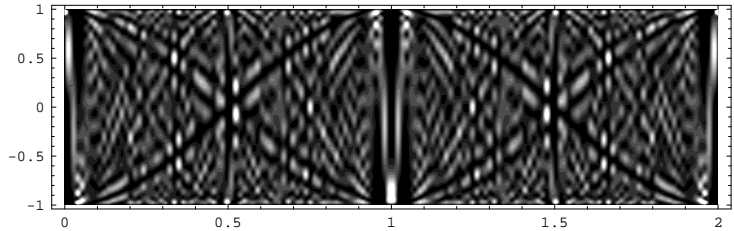}
\caption{The density plot of $|\tilde{\varphi}(x,t)|^2$ for,
$\g=10$ and the maximum value of $n=20$ in the symmetric
P{\"o}schl-Teller potential. Darkness displays a low and
brightness a high functional value.}
\end{figure}

\begin{figure}
\centering
\includegraphics[width=3in]{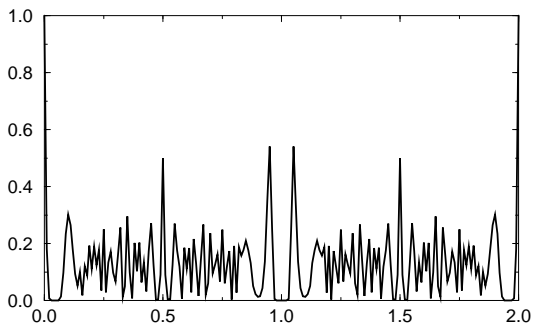}
\caption{Squared modulus $\mid_{\rm SPT}\la x,\g,t\neq 0 \mid
x,\g,t=0 \ra_{\rm SPT}\mid^2$ ($y$-axis) and time ($x$-axis), for
$\g=10$, $\r=2$, and $n=20$, of the autocorrelation for the SPT
potential. The peaks show the revivals whereas the intermediate
ones fractional revivals}
\end{figure}

\begin{figure}
\centering
\includegraphics[width=3in]{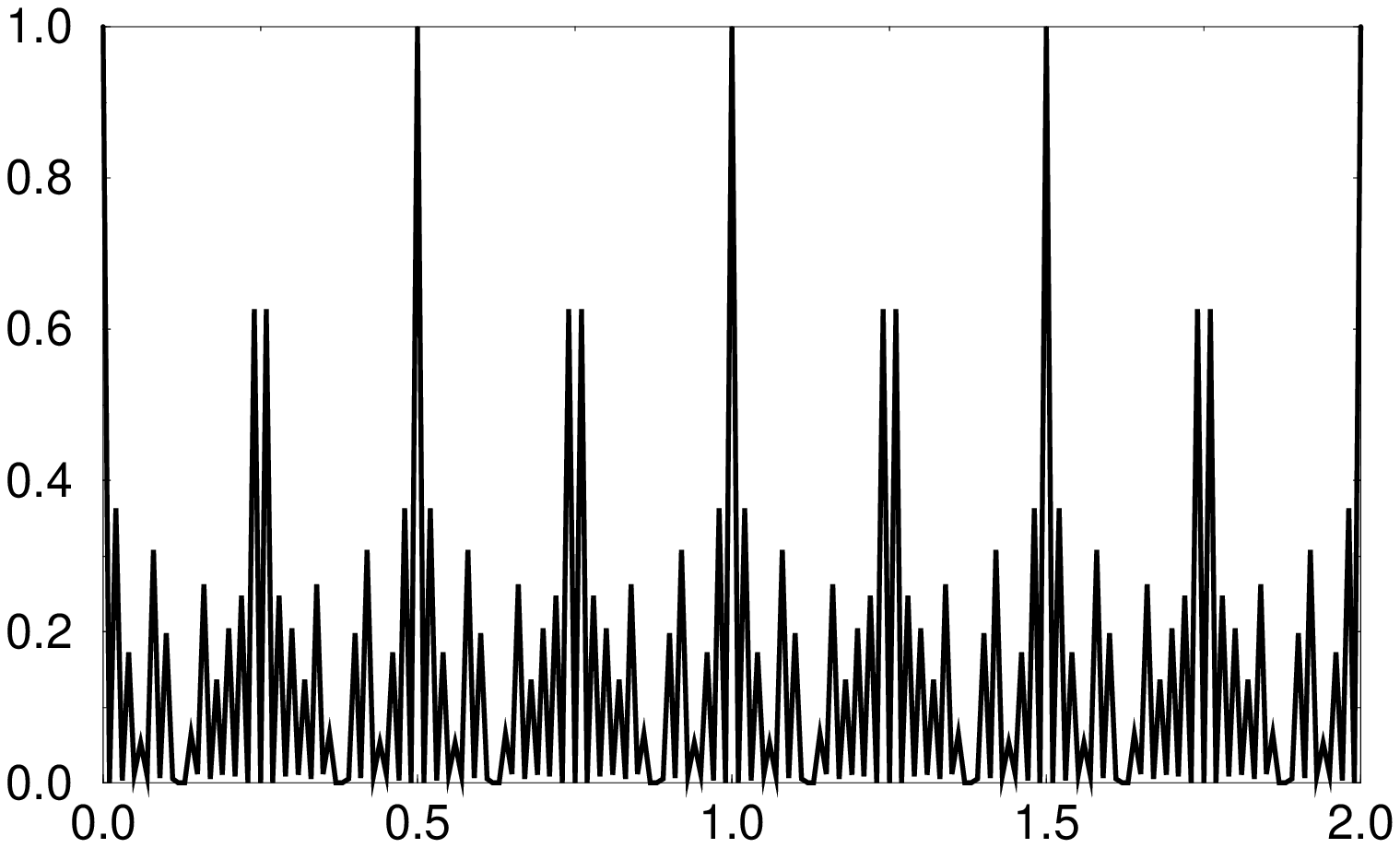}
\caption{Squared modulus $\mid_{\rm PT}\la x,\g,t\neq 0 \mid
x,\g,t=0 \ra_{\rm PT}\mid^2$ ($y$-axis) and time ($x$-axis), for
$\g=10$, $\k=2$, $\r=6$, and $n=20$, of the autocorrelation for
the PT potential.}
\end{figure}


As is well-known, the weight factors associated with the CS carry
physical significance e.g., these factors for the oscillator CS
give rise to a Poisson distribution. The weight factors associated
with the HG class of CS, derived above, are related to the HG
distribution in probability theory \cite{feller}, as is clear from
the plot of the weighting distribution in Fig (5).


\begin{figure}
\centering
\includegraphics[width=3in]{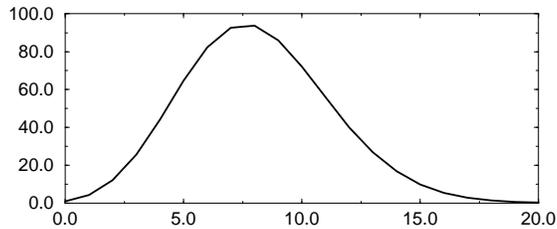}
\caption{Plot of the weighting distribution $|c_n|^2$ ($y$-axis)
in Eq. (\ref{mptcs}), with respect to $n$ ($x$-axis). This plot
captures the HG statistical nature of the coherent state. Here
$\g=10$, $\r=2$, and the maximum value of $n$ is 20.}
\end{figure}


\section{Conclusions}

In conclusion, we have developed a general algebraic procedure for
finding the annihilation operator coherent states for a wide class
of quantum mechanical potentials. Interestingly, the Perelomov
type CS also emerged, as AOCS, from different realizations of
relevant symmetry algebras. Crucial use was made of a simple
method for solving linear DEs, which gives a precise connection
between the solution space and the space of monomials. Ladder
operators, corresponding to various symmetry algebras can be
identified straightforwardly. The method is potential independent
and enables one to find the AOCS and connect them with earlier
oscillator based approaches. It is applicable to quantum problems
having infinite number of bound states as well as the ones
possessing finite numbers. Generalizations to analogs of the
nonlinear CS for oscillators is also made transparent.

Under time evolution, these CS showed revival and fractional
revivals. This manifests in the quantum carpet structure as well
as the auto-correlation functions. Interestingly, these phenomena
in the context of square-well \cite{frank} have led to a proposal
for factorizing numbers \cite{mack}. The intricate structure of
quantum carpet needs careful analysis in light of recent proposals
to use CS for quantum information storage \cite{enk}. The
weighting distributions associated with these CS also needs to be
studied more elaborately, in the complete parameter range, for
manifestation of non-classical behavior.

Also as a continuation of the present work, it would be
interesting to study the features of Wigner quasi-probability
distributions for these CS, in light of the interesting results
obtained recently in this area \cite{zurek}. It is worth noting
that, recently the Wigner distribution for the Morse eigenstates
have been studied \cite{wolf}. Since the method used here also
applies to many-body interacting systems it is worth constructing
and studying the corresponding CS \cite{chaturvedi}. A number of
these questions are currently under study and will be reported
elsewhere.

\end{document}